\documentclass[conference]{IEEEtran}
\IEEEoverridecommandlockouts
\UseRawInputEncoding
\usepackage{graphicx}
\usepackage{amsmath}
\usepackage{amsfonts}
\usepackage{mathrsfs}
\usepackage[OT1]{fontenc} 
\usepackage{enumerate}
\usepackage{amsthm}
\usepackage{tabularx}
\usepackage{arydshln}

\usepackage[colorlinks=true,linkcolor=blue,citecolor=blue]{hyperref}
\usepackage{mwe}
\usepackage{subfig}
\usepackage{float}
\usepackage{graphicx}
\usepackage{textcomp}
\usepackage{xcolor}
\usepackage{footnote}
\usepackage{makecell}
\usepackage{graphicx}
\usepackage{amsmath}
\usepackage{amsfonts}
\usepackage{mathrsfs}
\usepackage[OT1]{fontenc} 
\usepackage{enumerate}
\usepackage{amsthm}
\usepackage{tabularx}
\usepackage{arydshln}

\hypersetup{hidelinks}
\usepackage{subfig}
\usepackage{textcomp}
\usepackage{xcolor}
\usepackage{footnote}
\usepackage{makecell}
\usepackage[draft]{todonotes}

\usepackage{algorithm}% http://ctan.org/pkg/algorithms
\usepackage{algpseudocode}% http://ctan.org/pkg/algorithmicx
\usepackage[inline]{enumitem}
\usepackage{mathtools}

\definecolor{bleudefrance}{rgb}{0.19, 0.55, 0.91}
\definecolor{chromeyellow}{rgb}{1.0, 0.65, 0.0}
\definecolor{asparagus}{rgb}{0.53, 0.66, 0.42}

\begin{document}
\title{Assessing Connected Vehicle Data Coverage on New Jersey Roadways\\
\thanks{}}
%}

\author{
\IEEEauthorblockN{Branislav Dimitrijevic, Zijia Zhong, Liuhui Zhao, 
 Dejan Besenski, Joyoung Lee} %, \textit{member} IEEE
\IEEEauthorblockA{John A. Reif, Jr. Department of Civil and Environmental Engineering\\
\textit{New Jersey Institute of Technology}\\
Newark, NJ, USA \\
\{bxd1947, zijia.zhong, liuhui.zhao,  besenski, jo.y.lee\}@njit.edu}
%\and‼
%\IEEEauthorblockN{2\textsuperscript{nd} Liuhui Zhao}
%\IEEEauthorblockA{\textit{Department of Civil and Environmental Engineering} \\
%\textit{New Jersey Institute of Technology}\\
%Newark, NJ, USA \\
%liuhui.zhao@njit.edu}
}

\maketitle
\begin{abstract}
The connected vehicle data (CVD) is one of the most promising emerging mobility data that  greatly increases the ability to effectively monitor transportation system performance. A commercial vehicle trajectory dataset was evaluated for market penetration and coverage to establish whether it represents a sufficient sample of the vehicle volumes across the statewide roadway network of New Jersey. The dataset (officially named Wejo Vehicle Movement data) was compared to the vehicle volumes obtained from 46 weight-in-motion (WIM) traffic count stations during the corresponding two-month period. The observed market penetration rates of the Movement data for the interstate highways, non-interstate expressways, major arterials, and minor arterials are  2.55\% (std. dev. 0.76\%),  2.31\% (std. dev. 1.07\%),  3.25\% (standard deviation 1.48 \%), and 4.39\% (standard deviation 2.65\%), respectively. Additionally, the temporal resolution of the dataset (i.e., the time interval between consecutive Wejo vehicle trips captured at a given roadway section, time-of-day variation, day-of-month variation) was also found to be consistent among the evaluated WIM locations. Although relatively low (less than 5\%), the consistent market penetration, combined with uniform spatial distribution of equipped vehicles within the traffic flow, could enable or enhance a wide range of traffic analytics applications.

\end{abstract}


%%%%%%%%%%%%%%%%%%%%%%%%%%%%----------------------------INTRODUCTION-----------------------------------------%%%%%%%%%%%%%%%%%%%%%%%%%%%%%%%%%%%
% \todo[inline]{REVIEWER COMMENTS: \\
% 1. The captions of the all the tables should be consistently above the table.
% 2. It is highly recommended that the authors present a table of comparison of the proposed work with the research
% works studied as part of literature review. This will go a long way in emphasizing the research gaps and
% highlight ing the specific contributions of the proposed work.
% 3. It is recommended that the authors present a clear list of research gaps and a clear list of the contributions too.
% 4. All the first usages of abbreviations should be accompanied with the corresponding long forms. At the same
% location, the words of long form should be suitably in Title Case.
% 5. Some latest development coul d be considered. Additional ref –
% Parthasarathy Nadarajan, M
% ichael Botsch, and Sebastian Sardina, "Machine Learning Architectures for the Estimation
% of Predicted Occupancy Grids in Road Traffic," Vol. 9, No. 1, pp. 1 1-9, February 2018. doi: 10.12720/jait.9.1.1 9.1.1-9
% }

\section{Introduction}
%The observabilty of the transportation system is of paramount important in transportation engineering in terms of planning, operation \& maintenance, and incident management. 
%Sensors, which are used to collect traffic information, can be divided into stationary sensors (e.g., magnetic loop detector, camera) and mobile sensor (smart-phones, Bluetooth, aerial drone) [ref]. The mobile sensors landscape has shifted rapidly owing to the modern wireless connectivity and advancement in sensor technologies.

%Probe vehicle data
Probe vehicle data (PVD) is a mobile sensing method that has been widely used for ad-hoc traffic data collection. The scale of the PVD has evolved from the initial floating car runs using one or several probe vehicles, to fused data from embedded global navigation satellite system (GNSS) trackers and cellphone mobile applications used by the participating drivers. The PVD has been used in numerous applications. Considering their spatial resolution, the PVD can be grouped into two groups: segment-based PVD and point-based PVD. The former provide an aggregate (prevailing) or disaggregate travel time and space-mean speed of vehicles along predefined roadway segments (e.g., traffic message channel (TMC) links, or segments divided by the position Bluetooth sensors), based on the positions and timestamps of tracked vehicles as they enter and exit the corresponding roadway segments. The segment-based PVD have been adopted by numerous transportation agencies, researchers, and practitioners. The point-based PVD report the waypoints of individual vehicles that constitute a vehicles trajectory as it travels the roadway network. 

The PVD contain the data collected from a sample of vehicles in the traffic stream equipped with transponders that report the vehicle position at a certain preset frequency. Hence, the quality and reliability of analytics derived from the PVD depends on the size of vehicle sample relative to the total vehicle flow, which is referred to as "market penetration rate" (MPR). The GNSS-transponder-based PVD can be augmented (fused) with the vehicle positioning data from smart devices, which can substantially boost the MPR. However, such data is less consistent both in spatial and temporal respect due to the heterogeneity of the sensor pool \cite{pack2019considerations}. For instance, the use of low-power consumption mode of mobile devices is an inherent disadvantage for sourcing PVD. The INRIX Trip Analytics is an example of a point-based PVD that uses fused sources of vehicle trajectory data, and reportedly covers 10\% of all vehicle trips in the United States \cite{NREL2020Examines}. Other sources of point-based PVD typically have smaller MPR.

% However, the progress of adopting CAVs has been slow in recently years due to technical and other exogenous factors (e.g., global supply chain disruption caused by the COVID-19 pandemic).  

Even with its relatively sparse coverage at times (e.g., certain remote areas, night-time hours), the PVD have proven to provide substantial value for traffic analytics applications. Fan et al. compared 4-month INRIX trajectory data collected in 2015 against 84 continuous traffic count stations that were distributed throughout the State of Maryland. It was concluded that the PVD represented 1.5-2\% of the total traffic flow on Maryland roadways and offered a promising approach to estimating vehicle-miles traveled (VMT) \cite{fan2019using}.
The INRIX trajectory data was used to develop GNSS-based automated traffic signal performance measures (ATSPM) that was found to be spatially superior to segmented PVD (e.g., TMC travel time). The author also reported that an MP of the trajectory data in the range of 4\% -10\% was suitable for the proposed ATSPM, which requires a ping frequency of 5 s or less. Even with the MP below 0.04\% (less than 1\%), the proposed trajectory-based ATSPM allowed a quick diagnostics of immediate issues in a signalized corridor\cite{waddell2020utilizing, waddell2020characterizing}.

The emerging connected vehicles (CVs) with the built-in wireless connectivity are expected to greatly improve the collection of data about the transportation systems. Recently, a new form of commercialized probe vehicle data has emerged that is collected from vehicle fleets through exclusive agreements with automotive original equipment manufacturer (OEM), such as Wejo, Otonomo, High-Mobility. The scale of such commercialized CV data presents an immense potential for transportation planning, operation \& maintenance, and incident management. Besides high-resolution waypoint data,  these emerging data sets also provide vehicle telemetry data (e.g., fuel level, anti-lock braking system engagement, hard braking, windshield wiper activation). While "connected car data" has been used at times to differentiate this data from traditional PVD, these two terms are used synonymously in this paper. Based on Wejo's internal estimation, it receives vehicle telemetry data from 1 in 20 vehicle in the U.S., and 1 in 50 vehicles in Europe. Otonomo reportedly has more than 4 billion data points ingested into its platform daily from over 40 million licensed vehicles \cite{Otonomo}.  

Khadka et al. demonstrated an application of Wejo trajectory data for detection of queue propagation at freeway bottlenecks, as well as the assessment of traffic congestion on arterial roadways based on the percentage of slow moving vehicles \cite{khadka2022developing}.
For safety-related applications, Wejo harsh-braking event data was used as a safety surrogate measure in a before-and-after study of conversion of a left-turn phase from protected-permitted to protected. The same authors used Wejo event data in a different study and found a strong correlation between rear-end crashes and harsh-braking events \cite{hunter2021proactive}. In another study, the impact of a long-term work zone on trip diversions onto alternate routes (i.e., local arterial) was assessed using Wejo trajectory data. Owing to ubiquitous coverage of Wejo trajectory data, 100 intersections in the impact area of the work zone were evaluated during a 11-week period \cite{saldivar2021longitudinal}.

The market penetration of Wejo trajectory data in various areas has been reported. In the Dallas-Fort Worth area in Texas, the Wejo data presented 10\%-15\% of all moving vehicles \cite{khadka2022developing}. In the State of Indiana, the Wejo data was compared to traffic counts at 24 locations on Indiana roadways. The MPR was found to be 4.3\% with standard deviation (STDEV) of 1.0\% on interstate highways, and 5.0\% with STDEV of 1.36\% on non-interstate highways \cite{hunter2021estimation}.
A study that used Wejo trajectory data (collected in 2019) to estimate border crossing times, reported an insufficient MPR: around 30\% of the test hours were found to have no Wejo samples at the Paso del Norte port of entry in El Paso, TX. However, the study also confirmed the strong correlation between the Wejo travel time and the Bluetooth travel time within the border crossing area.\cite{li2021exploring}. 

The goal of this study is to assess the spatial-temporal coverage and consistency of Wejo probe vehicle data, which are critically important when considering application of vehicle trajectory data both in established and novel frameworks for assessing the performance of a transportation system. 
The highlights of this study are the following: 
\begin{enumerate*}[label=\roman*)]
\item the study scope focused exclusively on the New Jersey roadways across four roadway functional classifications that span all NJ counties, except Cape May County, \item the evaluation covers 46 permanent traffic count stations that provide hourly volume counts throughout a day, and \item the temporal analysis of the data is performed at various levels of aggregation.  
\end{enumerate*}

%%%%%%%%%%%%%%%%%%%%%%%%%%%%----------------------------METHODOLOGY-----------------------------------------%%%%%%%%%%%%%%%%%%%%%%%%%%%%%%%%%%%

\section{Evaluation Framework}

\subsection{Data}
The CV data used in the study (acquired from Wejo Group Ltd.) is exclusively crowd-sourced from automotive OEMs. It provides vehicle waypoints (latitude and longitude) and associated timetimestamps, along with other vehicle telemetry metrics reported at 1-3 second intervals. The trajectory coordinates are reported at 6-digit decimal precision, which is at the lane-level (i.e., 3-meter radius). The Wejo data used in this study (officially named Vehicle Movement data) included complete vehicle trajectory dataset collected within the State of New Jersey from April 15 through June 13, 2021. 

The volume of the data, given its granularity, sample size, and coverage, posed a challenge in terms of data storage, processing, visualization, and analytics. For instance, the Movement data collected in New Jersey in June 2021 amounts to 8.17 billion data points from 22.18 million journeys.
It should be noted that a unique journey ID is assigned each time the ignition is turned on in an equipped vehicle. As such, trip chaining could result in multiple unique journey IDs representing a single journey. It is assumed that within an hourly window, the vehicle traversing the bounding box is negligible. 

\begin{figure} [h]
	\centering
	\includegraphics[width=0.45\textwidth]{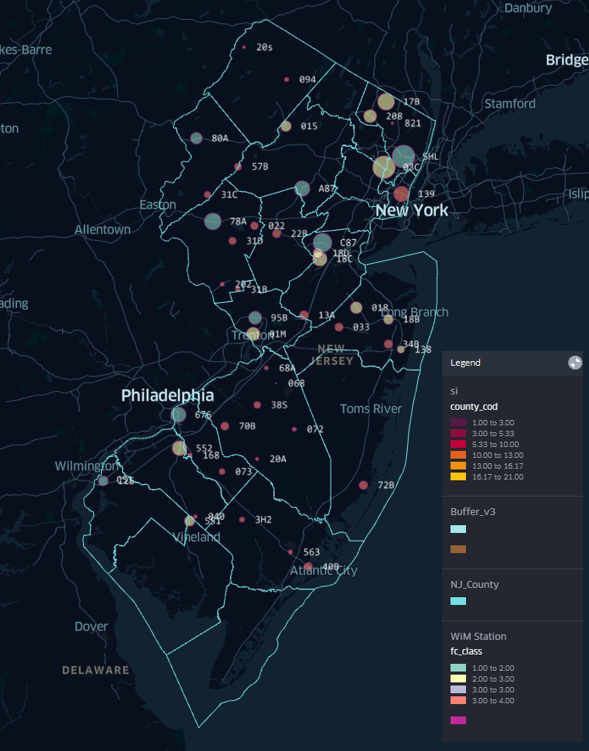}   
	\caption{46 WIM permanent count stations} 
	\label{fig:WiMStation}
\end{figure}

The number of unique journey IDs derived from the trajectory data is compared to the hourly traffic counts collected at 46 weigh-in-motion (WIM) stations throughout the State of New Jersey as shown in Fig. \ref{fig:WiMStation}. All of the WIM stations are classified as permanent count stations and are operated by the New Jersey Department of Transportation. The WIM data used in this study is publicly available via the NJTMS data portal \cite{NJTMS}.  As shown in Table \ref{table:FC_breakdown}, the 46 WIM stations cover the roadways classified in four functional classes, from freeways to minor arterials. 

\begin{table}[h]
\caption{Roadway Classification \& WIM Stations}
\resizebox{0.5\textwidth}{!}
{
\begin{tabular}{l|l|l}
\hline
\begin{tabular}[c]{@{}l@{}}Functional\\ classification\end{tabular}  & \begin{tabular}[c]{@{}l@{}}Roadway\\ category \end{tabular} & Number of stations \\ \hline
1 & Interstate & 8 \\ \hline
2 & other   freeway/expressway & 12 \\ \hline
3 & other   principle arterial & 20 \\ \hline
4 & minor   arterial & 6 \\ \hline
 & Total & 46 \\ \hline
\end{tabular}
}
\label{table:FC_breakdown}
\end{table}

To extract the vehicle waypoints from the trajectory dataset that correspond to each WIM station, the WIM station locations were geofenced by latitude/longitude boundaries covering approximately a 100 meter (300 ft.) buffer zone upstream and downstream. Manual verification was conducted to ensure the buffer zones do not include intersecting or adjecent roadways (e.g., underpass, overpass, frontage road), which could compromise the accuracy of the waypoint count corresponding to the WIM count. To process the massive amount of data in trajectory dataset, an Apache Spark clustering was employed.

\subsection{Metrics}
\subsubsection {MPR}
Market penetration is expressed as a percent of the entire (total) vehicle volume in a given area within a certain period contained in the trajectory dataset. In this study, the MPR for each hourly window was calculated using Eq. (\ref{eq:MPR})
\begin{equation}
\label{eq:MPR}
p^{ik}_{}=\frac{V'^{ik}_{m}}{V^{ik}_{}}
\end{equation}
where $p^{ik}_{}$ is the MPR observed during hour $k$ at count station $i$, $k \in \left \{ 0,1,...,23 \right \}$ and $i \in \left \{ 1,2,...,46 \right \}$; $V'^{ik}_{m}$ is the observed Wejo volume at station $i$ during hour $k$ and $V'^{ik}$ is the observed total volume at station $i$ during hour $k$.

As part of the data processing, some hourly intervals were excluded from the analysis. Specifically, an hourly interval was excluded if the corresponding WIM station had no count report for that hour, or had reported hourly volume of less than 50 vehicles. For WIM stations configured to collect and report two-way vehicle counts, the hours in which they reported only one-way counts were also excluded from the comparative analysis (assuming the sensor(s) were offline). Eventually, 52,496 hourly counts were included in the comparative analysis (with 3,206 hourly records removed). 

\subsubsection {Spatial Distribution}
The interval between the consecutive vehicles within the WIM station geofence was also analyzed for each WIM station. This concept is akin to time headway and can indirectly provide insights into the spacing between the tracked vehicles within the traffic stream. Assuming that their speed remains constant at a given roadway segment (e.g., equal to speed limit), the distribution of ping intervals is in essence equivalent to the distribution of time headway between the consecutive vehicles captured in the trajectory dataset. To remove the possible long ping intervals, the sample points during nighttime hours (from 9 PM to 5 AM) were excluded from the analysis.
% As an result, a total of 2,562 points were removed from consideration. 

\begin{equation}
\label{eq:HW}
h^{ik}_{n}=\frac{ \sum_{2}^{n} (t^{ik}_{n} -t^{ik}_{n-1})}{n}
\end{equation}
where $h^{ik}_{n}$ is the $nth$ time interval for station $i$ during hour $k$, $t^{ik}_{n}$ is the timestamp when the $nth$ vehicle entered the WIM station bounding box, and $t^{ik}_{n-1}$ is the timestamp when the $(n-1)th$ vehicle entered the WIM station bounding box, $n \geq 2$

%%%%%%%%%%%%%%%%%%%%%%%%%%%%----------------------------RESULT-----------------------------------------%%%%%%%%%%%%%%%%%%%%%%%%%%%%%%%%%%%

\section{Results}
\subsection{Statewide MPR}
\begin{figure} [h]
	\centering
	\includegraphics[width=0.5\textwidth]{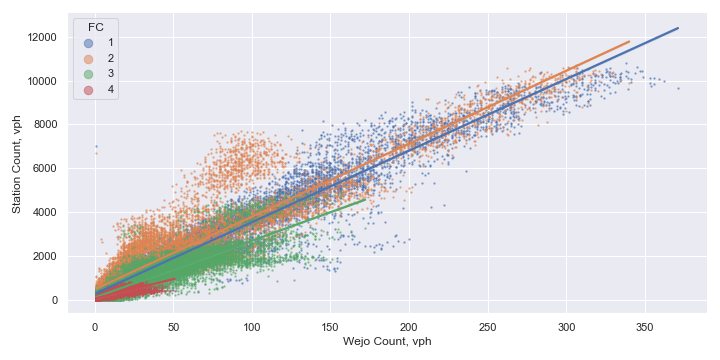}   
	\caption{Comparison of Wejo and hourly volume} 
	\label{fig:linearity_4cls_MPR}
\end{figure}

\begin{figure} [h]
	\centering
	\includegraphics[width=0.5\textwidth]{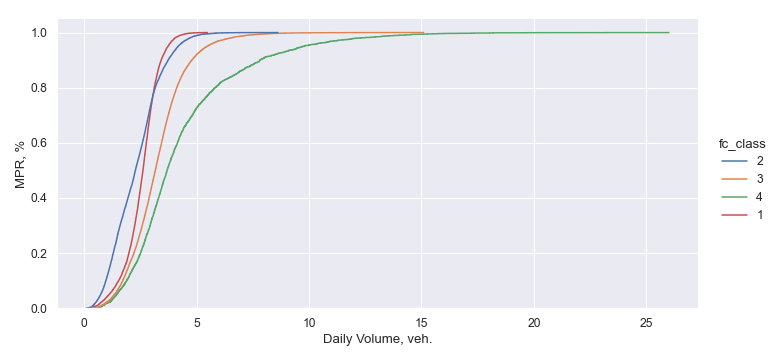}   
	\caption{Cumulative distribution curve for MP values} 
	\label{fig:cdf_mpr}
\end{figure}

The MPR statistics aggregated by roadway functional classification (FC) is summarized in Table \ref{table:mpr_results} and the linearity for each FC between the Wejo count and the WIM count is shown in Fig. \ref{fig:linearity_4cls_MPR}.  For FC=1 (Interstate highways), the MPR was 2.55\% with a STDEV of 0.75\%. The MPR for FC=2 (other freeways and expressways) was 2.31\% with a STDEV of 1.07\%. The MPR for FC=3 (other principal arterials) and FC=4 (minor arterials) are noticeably higher than those for FC=1 and FC=2: 3.25\% (Std. Dev 1.48\%) for FC=3, and the 4.39\% (STDEV 2.65\%) for FC=4. The low STDEV for FC=1 indicates that the representation of traffic stream by the vehicle trajectory dataset on interstate highways is relatively consistent. The data also shows that the MPR for FC=4 stations experienced higher variation as compared to the MPR at stations in other functional classes, which can be attributed to relatively low vehicle volumes recorded at these WIM stations (averaging 280 vehicles per hour). It is also of interest to note that the average MPR values are within the 2.5-4.5\% range, but their variations and probability distributions differ between different roadway functional classes. This can be seen in the cumulative distribution curves for each functional classification, shown in Fig. \ref{fig:cdf_mpr}. For example, the MPR for FC=1 caps at round 5.74\%, while the MPR for FC=3 was as high as 36.7\%. 

\begin{table}[h]
\caption{MPR Statistics}
\resizebox{0.5\textwidth}{!}
{
\begin{tabular}{l|l|l|l|l|l|l}
\hline
FC & Mean & Median & $r^2$ & Std. Dev. & Avg. Hourly Vol. & Avg. Daily Vol.  \\ \hline
1 & 2.55\% & 2.62\% & 0.949 & 0.76\% & 2,913 & 69,906  \\ \hline
2 & 2.31\% & 2.28\% & 0.829 & 1.07\% & 2,329 & 55,890  \\ \hline 
3 & 3.25\% & 3.15\% & 0.836 & 1.48\% & 853 & 20,465 \\ \hline
4 & 4.39\% & 3.69\% & 0.571 & 2.65\% & 282 & 6,763  \\ \hline
\end{tabular}
}
\label{table:mpr_results}

\end{table}

\begin{figure} [h]
	\centering
	\includegraphics[width=0.5\textwidth]{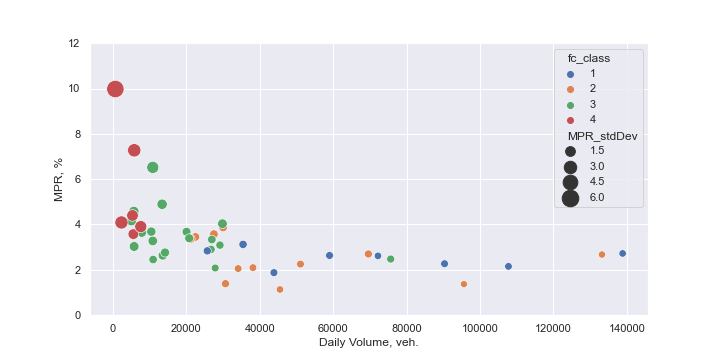}   
	\caption{MPR vs. station daily traffic count} 
	\label{fig:scatter_mpr-dailycount}
\end{figure}

Fig. \ref{fig:scatter_mpr-dailycount} shows the MPR pattern based on the daily vehicle volume recorded at the WIM stations across the roadway functional classes. With the circle size representing the values of standard deviations, one can observe that the higher the daily volume at a station, the lower the MPR and the standard deviation. For instance, the WIM stations at the FC=4 roadways tend to have lower daily volume and their MPR could be higher at times and typically have higher variance. For FC=1 and FC=2 stations, the MPR generally falls in the lower spectrum, but with smaller standard deviations.

\subsection{Spatial Distribution}
\begin{figure}[h]
\begin{minipage}[h]{0.5\textwidth}
\centering
\subfloat[FC-1(interstate highway), n=1,440]{\includegraphics[scale=0.3]{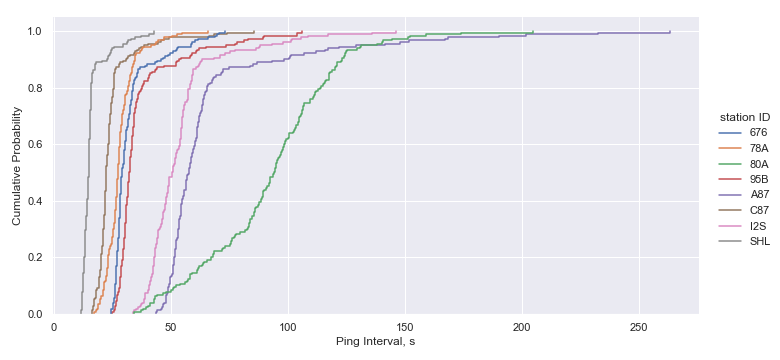}}
\end{minipage} \\
\begin{minipage}[h]{0.5\textwidth}
\centering
\subfloat[FC-2 (freeway/expressway), n=2,160]{\includegraphics[scale=0.3]{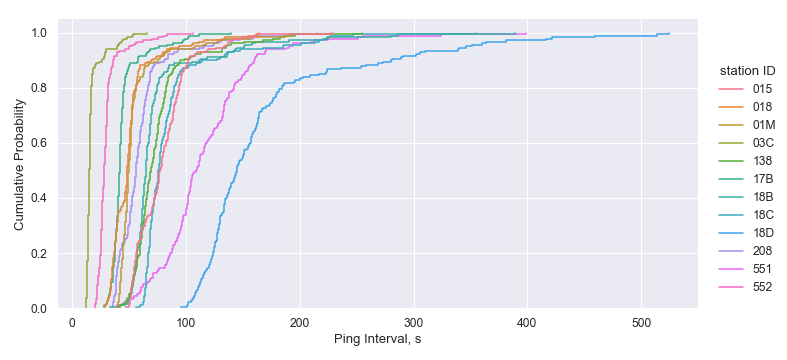}}
\end{minipage} \\
\begin{minipage}[h]{.5\textwidth}
\centering
\subfloat[FC-3 (principle arterial), n=3,958]{\includegraphics[scale=0.3]{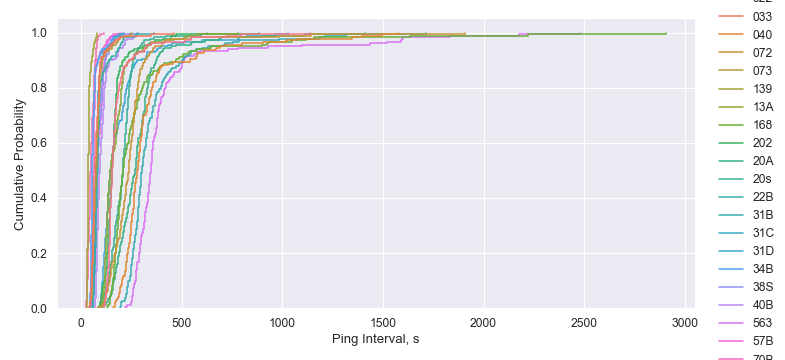}}
\end{minipage}
\caption{Equipped vehicle in traffic stream}
\label{fig:HW}
\end{figure}

The cumulative distribution curves of headways between the probe vehicles (i.e., the ping intervals) are plotted in Figs. \ref{fig:HW}(a), \ref{fig:HW}(b), and \ref{fig:HW}(c) for WIM stations located at roadways of functional classes FC=1, FC=2, and FC=3, respectively. The WIM stations at FC=1 roadways, with the exception of station ``80A'', all have more than 80\% of the ping intervals that are less than 60 seconds (some stations even have more than 80\% of the headways that are less than 40 seconds). Based on the analysis it can be concluded that the spatial distribution of the trajectory data is sufficiently representative of the overall traffic stream during daytime. Due to relatively long ping intervals and lower overall volume during nighttime, the trajectory data may not provide sufficient data points for conclusive analytics during nighttime hours.

\subsection{Temporal Variation}
\begin{figure} [h]
	\centering
	\includegraphics[width=0.5\textwidth]{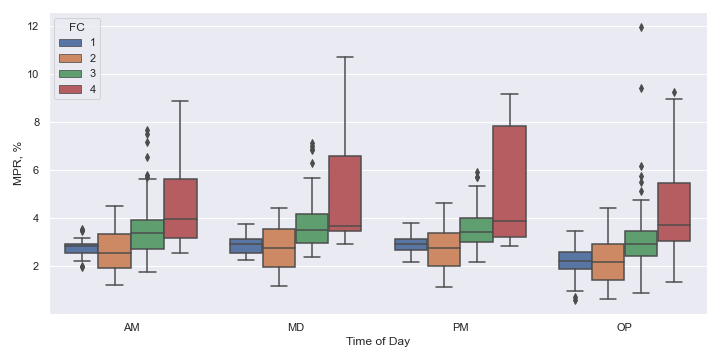}   
	\caption{MPR comparison for time of day} 
	\label{fig:mpr_tod}
\end{figure}

To investigate the time-of-day (ToD) variation, the MPR statics are grouped into four periods: ``AM-peak'' (6AM-10AM),  ``mid-day'' (10AM-3PM), ``PM-peak'' (3PM-7PM) , and``off-peak'' (7PM-6AM).  The MPR comparison among the four ToD windows is shown in Fig. \ref{fig:mpr_tod}. The roadways in FC=1 have the narrowest interquartile range (IQR) among the four classes, followed by FC=3, then FC=2. The WIM locations at roadways in FC=4 have the largest IQR in each ToD window and the MRP tends to skew towards the higher values. The median MPR tends to edge higher during daytime (6AM to 7PM) at all locations.

\begin{figure} [h]
	\centering
	\includegraphics[width=0.5\textwidth]{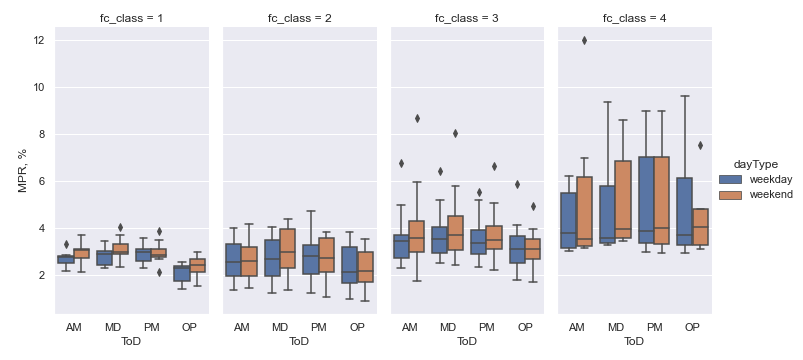}   
	\caption{Weekday-weekend comparison} 
	\label{fig:mpr_tod_weekend}
\end{figure}

The comparison of MPR during weekdays and weekends is shown in Fig. \ref{fig:mpr_tod_weekend}. We observe that the MPR values are slight higher during weekends for FC=1. This could be attributed to the optional trips (non-commute trips) on interstate highways. It was observed that the weekends typically had sample points with higher MPR values, as high as 9\%.

\begin{figure}[h]
\begin{minipage}[h]{0.5\textwidth}
\centering
\subfloat[Daily variation]{\includegraphics[scale=0.35]{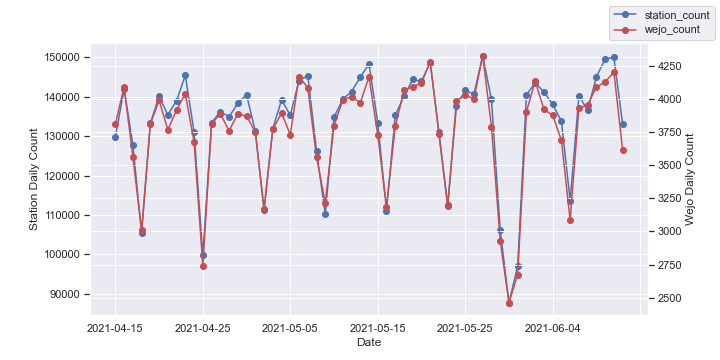}}
\end{minipage} \\
\begin{minipage}[h]{.5\textwidth}
\centering
\subfloat[Time of day variation]{\includegraphics[scale=0.35]{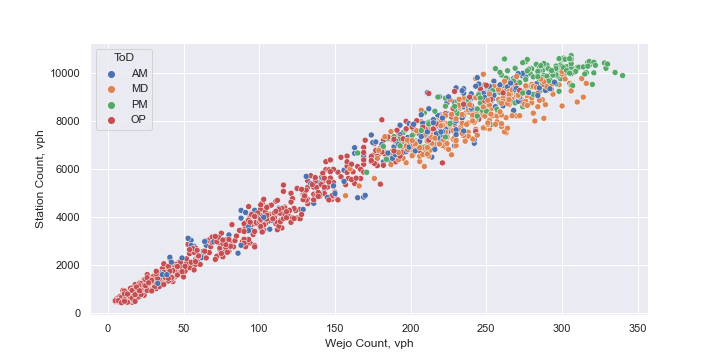}}
\end{minipage}
\caption{Variation (station ``03C'' on NJ-3, $r^2=0.977$)}
\label{fig:Stn03C}
\end{figure}

The following analysis also shows that the trajectory data, though being only a subset of the overall traffic stream, reflects well the time-of-day and day-of-month traffic patterns. The observation from WIM station ``03C'' (NJ Route 3 in Clifton Township) is presented as an example of high correlation between the WIM counts and the corresponding trajectory counts ($r^2=0.977$). Fig  \ref{fig:Stn03C} (a) shows the daily volume observed at the WIM station (left axis) and the Wejo count (right axis). Even though the two axes are of different scale, one can clearly see the Wejo volume closely tracking the daily variation of the ground-truth (WIM counts).  

\begin{figure}[h]
\begin{minipage}[h]{0.5\textwidth}
\centering
\subfloat[Daily variation]{\includegraphics[scale=0.35]{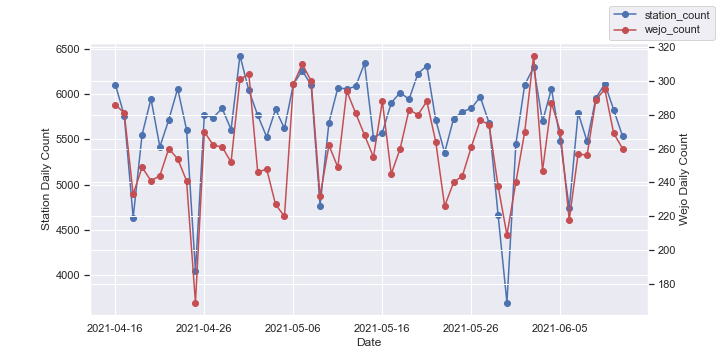}}
\end{minipage} \\
\begin{minipage}[h]{.5\textwidth}
\centering
\subfloat[Time of day variation]{\includegraphics[scale=0.35]{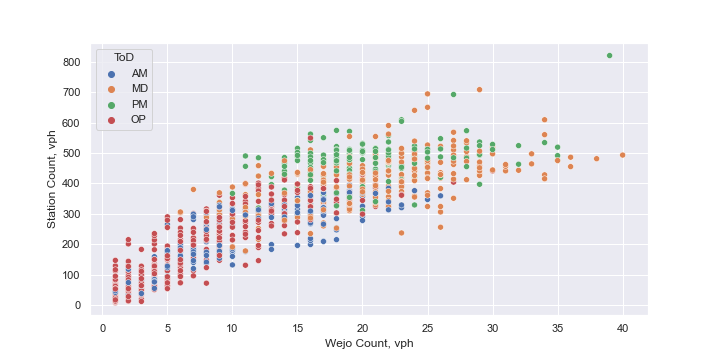}}
\end{minipage}
\caption{Variation (station ``168'' on NJ-168, $r^2=0.735$)}
\label{fig:Stn168}
\end{figure}

The ToD analysis for station ``03C'' is shown in  Fig. \ref{fig:Stn03C}(b). As expected, each period has the volume corresponding to the typical ToD variation with high traffic volume observed mostly in the AM and the PM peaks. Interestingly, both counts exhibit a high linear correlation. In another words, regardless the ToD, the MPR of the Wejo counts are highly consistent. It shoudl be noted that the WIM station ``03C'' is somewhat of an ideal case. For comparison, another WIM station is showcased, located on NJ-168, with less daily traffic. The daily WIM counts and trajectory counts at this station are less correlated ($r^2$=0.735), and the daily and ToD patterns are less aligned, as shown in Fig. \ref{fig:Stn168}.

\section{Discussion}
%discussion of potential applications

Based on the comparative analysis of vehicle counts and Wejo trajectory counts at 46 WIM stations across New Jersey, the spatial-temporal coverage and resolution of the trajectory data, make this data a great candidate for numerous traffic analytics applications. It was also observed that the trajectory data reflects traffic patterns even when the underlying MPR is less than 5\%.

Previous studies reported that the critical value of MPR varies by the types of applications. In another words, the trajecotry data with low MPR could still contribute to specific operational analysis. For instance, in the case of signal performance measures, 1\% penetration rate could provide insight into cyclic flow profiles of an intersection and similar performance could be achieved with the MPR as low as 0.1\% when PVD is stacked over multiple days \cite{day2016detector}. The signal coordination issues can be identified with less than 0.04\% MPR \cite{waddell2020characterizing}. For estimating vehicle miles traveled (VMT), 1.5-2\% MPR was found to be sufficient \cite{fan2019using}. Using the trajectory data is a promising and cost-effective approach for the volume and VMT estimation at roadways where traffic count data may be less reliable or infrequently collected, due to the ubiquitously coverage of Wejo data.

A study using Wejo data concluded that, based on the analysis of 8 intersections, one month of harsh-braking data is adequate for a reliable correlation with over 4-5 years' worth of crash data \cite{hunter2021proactive}. In certain cases, using the Wejo data for traffic analytics may present a challenge. For instance, a study of using Wejo data (2019) to estimate border crossing times found that the underlying Wejo MPR was insufficient, as 30\% of test hours lacked Wejo samples \cite{li2021exploring}.
It is our assessment, although this should be verified for specific locations and time periods, that the Wejo data generally reflects the actual traffic patterns, and could even be feasible source for real-time applications in active traffic monitoring and management practices, owing to the high resolution and consistent spatial and temporal representation of the traffic stream. 

\section{Conclusion}
\label{sect:conclusion}
This study conducted a large-scale evaluation of temporal-spatial converge and consistency of high-resolution commercial connected vehicle trajectory data (acquired from Wejo Group Ltd) on the New Jersey roadways. 17 billion records  of trajectory points (approximately 9 TB, uncompressed) were analyzed to extract the volume data for 46 permanent (WIM) count stations across New Jersey. The results show that the sample of the trajectory probe vehicles relative to the total vehicle volume (referred to as market penetration or representation) resides within the range of 2.4-4.5\%, on average, during daytime hours. In addition, the reviewed previous studies demonstrated general uniformity of the distribution of equipped vehicles in the traffic stream over various roadways, which could greatly increase the practicality of the trajectory data. Analysis of the temporal data variation (e.g., time-of-day, weekday-weekend comparisons) also shows the good representation of the traffic stream in general. The traffic analytics that can immediately benefit from the vehicle trajectory data include roadway risk profiling, crash timeline reconstruction, traffic incident impact monitoring, and driving behavior analysis.

% Our future research will focus on the prototype applications that make the full use of the high-resolution trajectory data. Furthermore, the near ubiquitous coverage allow use to relax assumptions in established models or framework, such as GNSS-based automated traffic signal performance measures.

%\section*{Acknowledgment}
%Provide acknowledgement here...
   
%\begin{enumerate*}[label=\roman*)]
%\item to test the effectiveness of CAV lane with clustering strategy for CAVs where a free-agent CAV actively seeks opportunity to form vehicle platoons,
%\item to investigate the potential impact to non-CAVs (HVs) due to induced lane change activity in the presence of CAV lane, which is typically local in the leftmost travel lane,
%\item to increase the understand of the impact of CAV at individual trajectory level by analyzing high-resolution vehicle trajectory data collected in simulation, and
%\item to implement more sophisticated CACC algorithm 
%\end{enumerate*} 

%\section*{Acknowledgment}
%
%This work is supported in part by the National Science Foundation under Grant No. CMMI-1844238.

\bibliographystyle{IEEEtran}
\bibliography{Wejo-MPR}
\end{document}